\documentstyle[11pt,aaspp4,psfig]{article}
\lefthead{Aparicio et al.}
\righthead{DDO 190}

\begin{document}

\title{The spatial and age distribution of stellar populations in DDO 190\footnote{Based on observations made with the 2.5 m Nordic Optical Telescope
operated on the island of La Palma by NOT S.A. at the 
Spanish Observatorio del Roque de Los Muchachos of the 
Instituto de Astrof\'\i sica de Canarias.}}

\author{A. Aparicio}
\affil{Instituto de Astrof\'\i sica de Canarias, E38200 - La Laguna,
Tenerife, Canary Islands, Spain}
\author{N. Tikhonov}
\affil{Special Astrophysical Observatory, Stavropol, Russia}

\begin{abstract}

The spatial distribution of stellar populations, the star formation history,
and other properties of the dIrr galaxy DDO 190 have been analyzed using
color--magnitude diagrams (CMDs) of about 3900 resolved stars and the
H$_\alpha$ fluxes of H~{\sc ii} regions. From the mean color index of the red
giant branch, a mean metallicity [Fe/H] $=-2.0$ is obtained. The $I$
magnitude of the TRGB has been used to estimate the distance. DDO 190 is 
$2.9\pm 0.2$ Mpc from the Milky Way, 2.1 Mpc from the M 94 group (CnV-I), 2.4
Mpc from the M 81 group and 2.9 Mpc from the barycenter of the Local Group,
all indicating that it is an isolated, field galaxy.

The surface-brightness distribution of the galaxy is well fitted by ellipses
of ellipticity $e=1-a/b=0.1$ and  P.A. $ = 82^\circ$. The radial star density
distribution follows an exponential law of scale length $\alpha=43\farcs4$,
corresponding to 611 pc. The Holmberg semi-major axis to $\mu_B=26.5$ is
estimated to be $r^B_{26.5}=3\farcm0$.

Stellar populations of different ages in DDO 190 show  strong spatial
decoupling, the oldest population appearing much more extended than the
youngest. Stars younger than 0.1 Gyr occupy only the central $40''$ (0.55
kpc); stars younger than a few ($\sim 4$) Gyr extend out to $\sim 80''$ (125
kpc), and for larger galactocentric distances only older stars seem to be
present. This behavior is found in all the dIrr galaxies for which 
spatially extended studies have been performed and could be related with the
kinematical history of the galaxy.

\end{abstract}

\section{Introduction}

Dwarf galaxies are usually divided in two main categories: dwarf irregular
(dIrr) and dwarf spheroidal/dwarf elliptical (dSph/dE). The most obvious
differences between them are internal and refer to the amount of gas and
young stellar populations. However dSph/dE and dIrr galaxies also show a
striking difference in their spatial distribution: while dSphs/dEs appear
concentrated around bigger galaxies or in the high-density regions of galaxy
clusters, dIrrs show a smooth distribution, with only 50 \% associated with
clusters (Karachentseva \& Vavilova 1994). This suggests that dIrrs evolve
into dSphs/dEs through gas-loss produced by external ram pressure and tidal
stripping, which would be more efficient in high-density environments and the
surroundings of big galaxies (Einasto et al. 1974; Lin \& Faber 1983).

Although suggestive, this picture needs to be tested on the basis of wider
and deeper observational information. Much work has been carried on in the
last few years analyzing dwarf galaxies in the Local Group (see the review by
Mateo 1998). But in order to have a complete overview, this kind of analysis is
also necessary for galaxies in the outskirts of the Local Group and in the
field. In this paper, the structural properties and the stellar population of
one of these objects, DDO 190, are studied on the basis of $V$, $I$, and
H$_\alpha$ photometry. 

DDO 190 (UGC 9240; $\alpha_{00}=14^{\rm h}24^{\rm m}43^{\rm s}$;
$\delta_{00}=+44^\circ31\farcm5$) is a resolved, gas-rich dwarf galaxy first
cataloged by van den Bergh (1959). It lies in a region of the sky with
a reddening of 
almost 0 (Burstein \& Heiles 1982). Its magnitude and colors have
been given by de Vaucouleurs et al. (1991: $B_T^0=13.35$, $(B-V)_0=0.35$,
$(U-B)_0=-0.17$), by Melisse \& Israel (1994: $B_T^0=13.1$, $(B-V)_0=0.24$,
$(U-B)_0=-0.21$) and by Prugniel \& H\'eraudeau (1998: $B_T^0=13.00$,
$(B-V)_0=0.29$, $(U-B)_0=-0.20$). Fisher \& Tully (1981) and Hutchmeier \&
Richter (1988) provide H~{\sc i} data for the galaxy which include integrated
fluxes and velocity dispersion. They obtain heliocentric velocities of
$v_\odot=153$ km s$^{-1}$ and $v_\odot=158$ km s$^{-1}$, respectively, and
relative to the Local Group velocities of $v_{\rm LG}=269$ km s$^{-1}$ and
$v_{\rm LG}=281$ km s$^{-1}$, respectively. Far-infrared data are given by Thuan
\& Sauvage (1992), by Isobe \& Feigelson (1992) and by Melisse \& Israel
(1994). Finally, upper limits to the CO emission have been obtained by Israel,
Tacconi, \& Baas (1995). Its distance has been never directly
estimated. Values in the range 2.8--5.6 Mpc have been frequently used, as
derived from $v_{\rm LG}$ and different choices of the Hubble parameter.

The paper is organized as follows: in \S 2 the observations and data
reduction are described. The color--magnitude diagram (CMD) of the resolved
stars is presented in \S 3. In \S 4 the distance and the metallicity are
derived from the magnitude of the tip of the red giant branch (TRGB) and the
color of the red giant branch (RGB). The distribution of ionized H is
discussed in \S 5. Section 6 is devoted to the study of the structure and
the spatial distribution of stars. The star formation history (SFH) has been
obtained using synthetic CMDs, which is presented in \S 7. Section 8
summarizes the global properties of DDO 190. Finally, in
\S 9 the main conclusions of the paper are given.

\section{Observations and data reduction}

\begin{figure}
\centerline{\psfig{figure=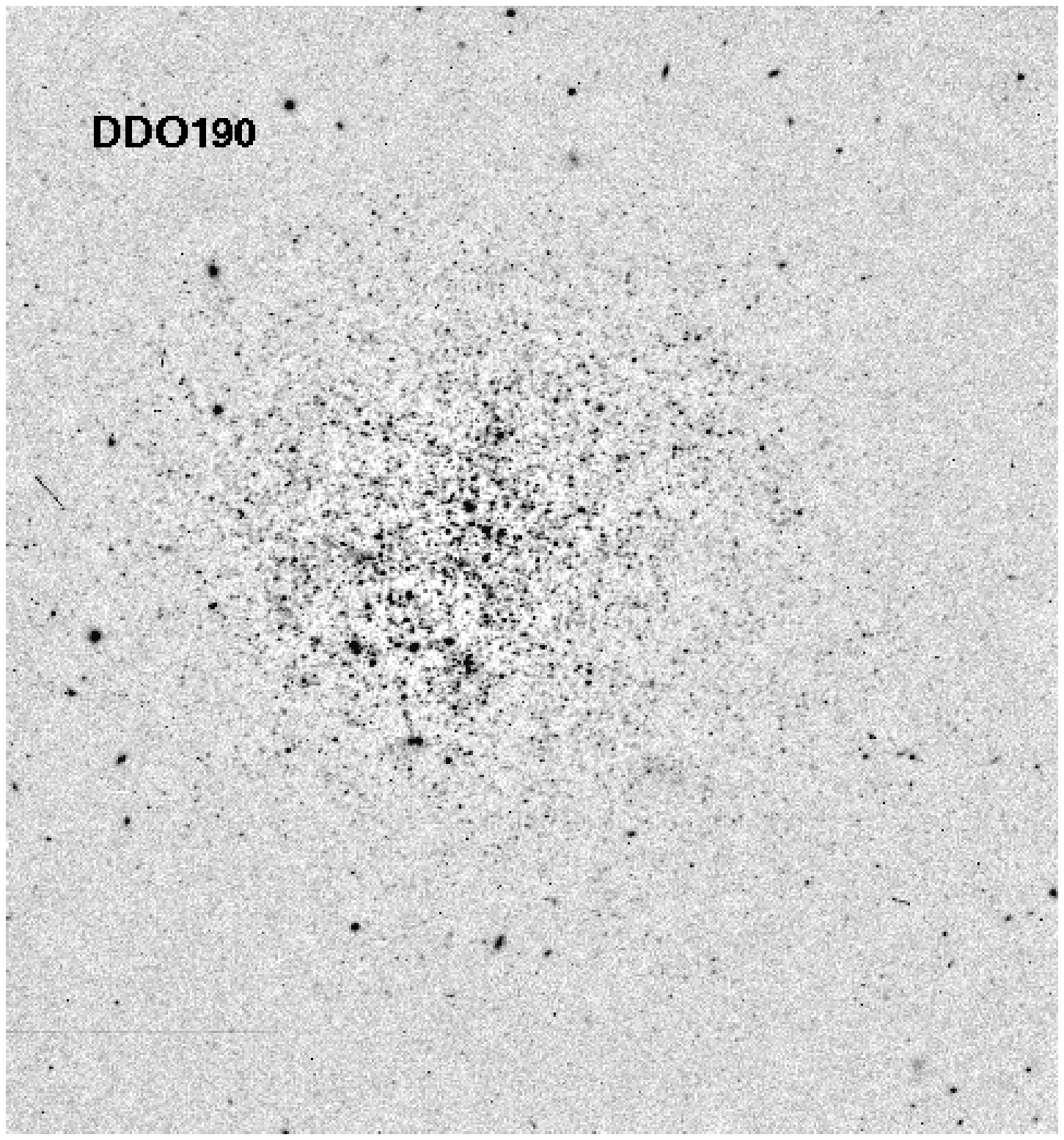,width=16cm}}
\figcaption[ima.eps]{$I$ image of DDO 190. The total field is
$3.75\times3.75(')^2$ and the integration time 1000 s. North is up, east
is to the left.
\label{ima}}
\end{figure}

Images of DDO 190 were obtained in the $V$ and $I$ Johnson--Cousins
filters and in a narrow band H$_\alpha$ filter with the NOT (2.5 m) at
Roque de los Muchachos Observatory on La Palma (Canary Islands,
Spain). The HiRAC camera was used with a $2048\times 2048$ Loral CCD binned to
$2\times2$. After binning it provides a scale of 0.22 $''$/pix and a total
field of $3.75\times 3.75$ $(')^2$. Total integration times were 2400 s in
$V$, 2000 s in $I$, and 1800 s in H$_\alpha$. A 600 s exposure was also
taken with an H$_\alpha$-continuum filter. Moreover, integrations of
1200 s in $V$ and 1200 s in $I$ were performed for a nearby field to
correct the foreground contamination of the broad-band images. Table \ref{journal}
gives the journal of observations. 
Figure \ref{ima} shows one of the $I$-band images of DDO 190. 

\placetable{journal}
\placefigure{ima}

Bias and flat-field corrections were carried out with IRAF. Then, DAOPHOT and
ALLSTAR (Stetson 1994) were used to obtain the instrumental photometry of the
stars. Eighteen standard stars from the list of Landolt (1992) were measured
during the observing run to calculate the atmospheric extinctions for each
night and the equations transforming into the Johnson--Cousins standard
photometric system. A total of about 180 measurements in  $V$ and $I$ of
these standards were used. The transformation equations are:

\begin{equation}
(V-v)=25.205-0.106(V-I); ~~~~\sigma=0.005
\end{equation}
\begin{equation}
(I-i)=24.498+0.011(V-I); ~~~~\sigma=0.004
\end{equation}

\noindent where capital letters stand for Johnson--Cousins magnitudes and
lower-case letters refer to instrumental magnitudes corrected for
atmospheric extinction. The $\sigma$ values are the dispersions of the
fits at the barycenters of the point distributions;
hence they are the minimum internal zero-point errors. Dispersions of the
extinction for each night varies from $\sigma=0.009$ to $\sigma=0.019$
and the dispersion of the aperture corrections are of the order of
0.04. Putting all these values together, the total zero-point error of
our photometry can be estimated at about 0.05 for both bands.

ALLFRAME provides PSF-to-star fitting errors. These are 0.02, 0.05 and 0.2
for $I=20.0$, 22.0 and 23.8, respectively. However, these are not the real
external errors of the photometry, which are bigger and can be properly
obtained only through detailed artificial star trials (Aparicio, \& Gallart
1995).

One standard star from Oke (1990) was observed three times to calibrate the
H$_\alpha$ images. The dispersion of the measures was better than 0.01
magnitudes.

\section{The color--magnitude diagram}

\begin{figure}
\centerline{\psfig{figure=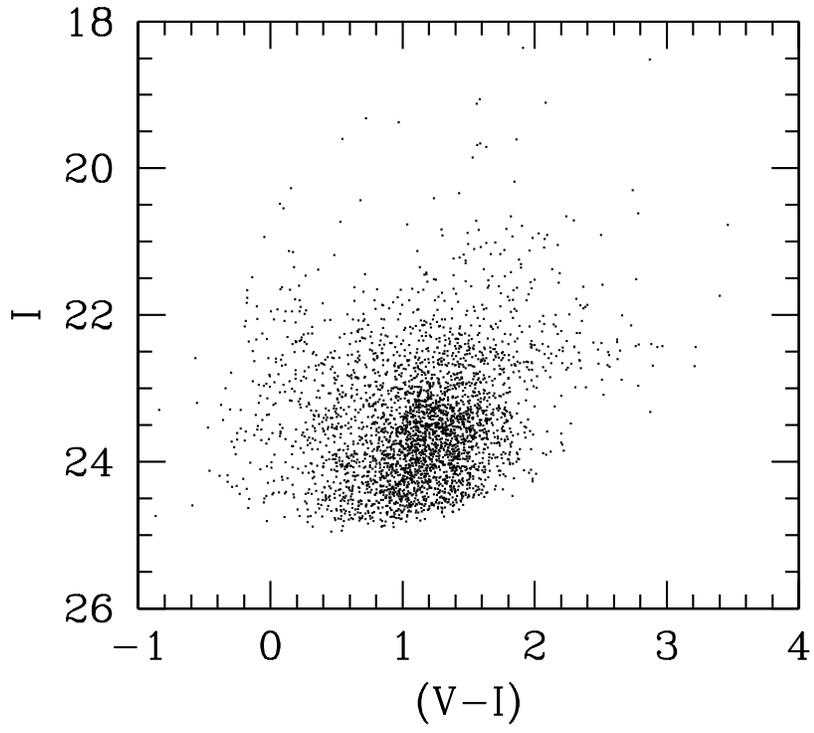,width=16cm}}
\figcaption[cmd.eps]{CMD of DDO 190.
\label{cmd}}
\end{figure}

\placefigure{cmd}

The CMD of DDO 190, shown in Figure \ref{cmd}, is typical of a resolved
galaxy with recent star formation. It shows similar structures as that of DDO
187 (Aparicio, Tikhonov, \& Karachentsev 2000), but with a less well defined
appearance. A well populated structure is visible in the region $I\geq 23$,
$0.8\leq (V-I)\leq 1.8$, corresponding to the {\it red-tangle} (see Aparicio
\& Gallart 1994; Aparicio et al.  1996). It is formed by low-mass stars
in the RGB and AGB phases and it is hence the trace of an intermediate-age to
old stellar population. It shows a large color dispersion. Although this may
be related with a large metallicity dispersion, it is more likely due to
external photometrical errors, which we have not analyzed here in detail.

The CMD shows also a number of blue stars ($V-I\leq
0.6$). Some of these may be foreground stars, but many are probably members of
DDO 190, with intermediate to high masses in the main sequence
(MS) and/or the core He-burning blue-loop evolutionary phases. They are the
trace of a significant recent star formation, in good agreement with the
presence of H~{\sc ii} regions in the galaxy. Finally, some of the red stars in the
region $20.5\leq I\leq 23$, $(V-I)\geq 1.1$ are probably intermediate-age to
young AGB stars. They also form a {\it red-tail} structure of the kind discussed by
Aparicio \& Gallart (1994), visible at $I\simeq 22.5$, $(V-I)\geq 1.7$.

\begin{figure}
\centerline{\psfig{figure=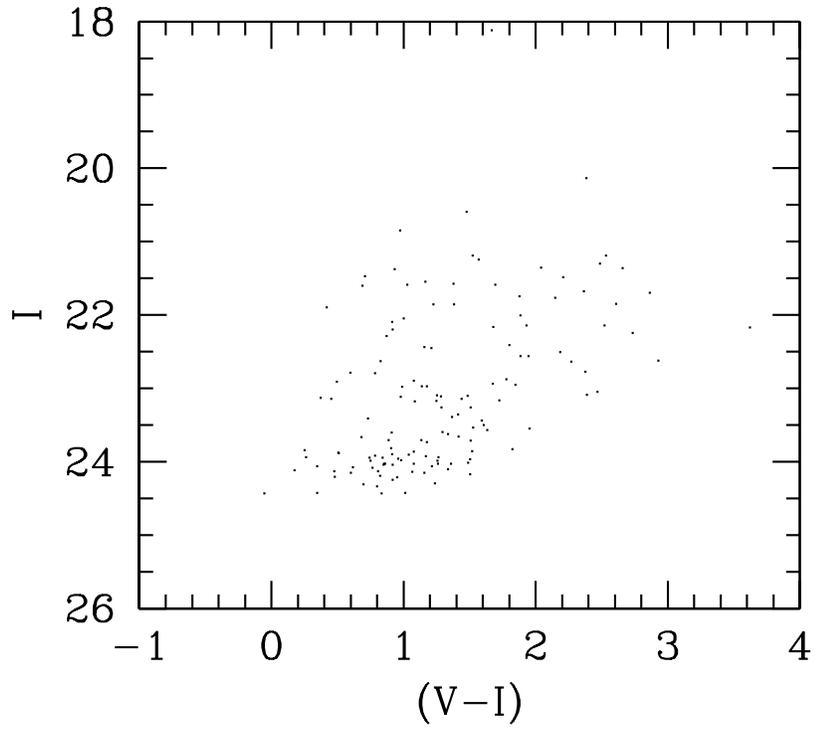,width=16cm}}
\figcaption[cmd_f.eps]{CMD of a DDO 190 nearby companion field to map the
foreground contamination.
\label{cmd_f}}
\end{figure}

\placefigure{cmd_f}

Figure \ref{cmd_f} shows the CMD of a companion field, close to DDO 190 but
far enough to guarantee that it contains no stars of the galaxy. The
distribution of stars in this diagram indicates that some of the red and
yellow stars in the CMD of Figure   \ref{cmd} are foreground stars, belonging to
the Milky Way, but that most of the bright blue ones are probably members of
DDO 190.

In summary, the CMD of DDO 190 indicates that it has
had an extended SFH, as evidenced by the presence of young blue stars
and a red-tangle of intermediate-age and/or old stars. A quantitative
estimate of the SFH is given in \S 7.

\section{The distance of DDO 190}

\begin{figure}
\centerline{\psfig{figure=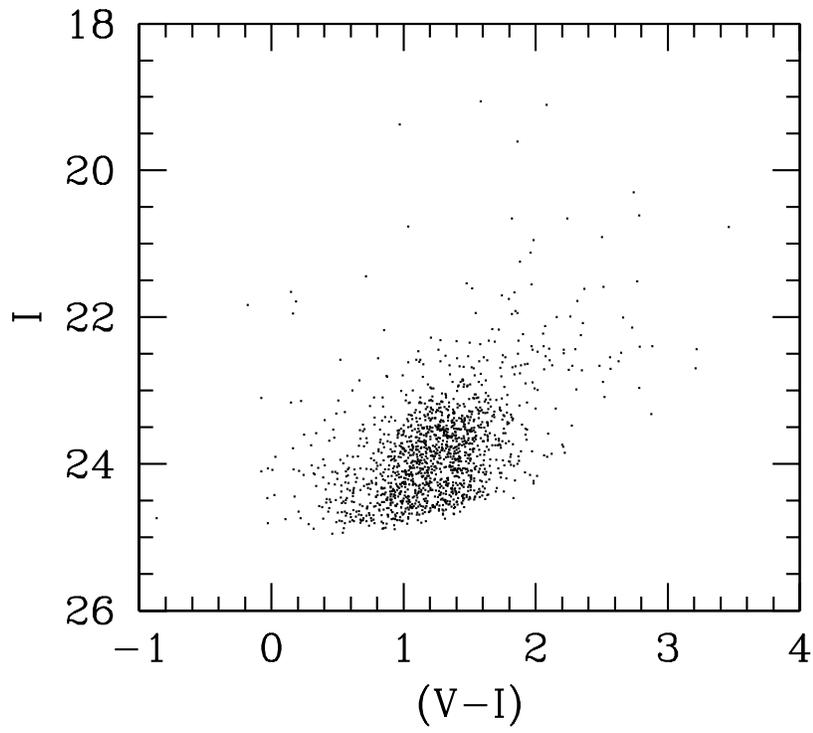,width=16cm}}
\figcaption[cmd_1arcmin.eps]{CMD of DDO 190. Only stars at more than $1'$
from the galaxy center have been plotted. This CMD has been used to estimate
the distance of DDO 190.
\label{cmd_1arcmin}}
\end{figure}

The distance to DDO 190 can be estimated using the magnitude of the TRGB and
the relationships by Lee, Freedman, \& Madore (1993). We have determined the
magnitude of the TRGB applying a Sobel
filter to the luminosity function (LF) of stars with colors in the interval
$1.2\leq (V-I)\leq 1.6$. To minimize crowding and blending effects, only
stars at distances greater than $1'$ from the center of the galaxy have been
used. The corresponding CMD is shown in Fig. \ref{cmd_1arcmin}. We have
obtained $I_{\rm TRGB}=23.32\pm 0.15$. The error has been estimated from the
width of the peak produced by the Sobel filter around the TRGB. Several
binning sizes and initial values for the LF have been used, the former result
being the average of all them. The extinction towards DDO 190 can be
neglected (Burstein \& Heiles 1982).

\placefigure{cmd_1arcmin}

To apply the relationships of Lee  et al. (1993), the metallicity of the
stars and the bolometric correction (BC) at the TRGB are necessary. Both can
be estimated from the color indices of RGB stars. $(V-I)_{\rm TRGB}=1.34\pm
0.03$ and $(V-I)_{-3.5}=1.27\pm 0.02$ have been obtained as the median of the
color indices of stars at distances greater than $1'$ form the center of the
galaxy, having colors in the interval $1.0\leq (V-I)\leq 1.6$ and with
magnitudes in the ranges $23.32\leq I\leq 23.42$ and $23.77\leq I\leq 23.87$,
respectively. Some 50 stars have been used in each case and the quoted errors
have been simply obtained as $\sigma\times (n-1)^{-1/2}$ for each
sample. From $(V-I)_{-3.5}$, an average metallicity of [Fe/H] $=-2.00\pm
0.08$ is found using the calibration by Lee et al. (1993). This Fe abundance
corresponds to $Z=0.00020\pm 0.00005$. It must be noted, however, that the
Lee et al.'s relation applies for old stellar populations. A
significant population of intermediate-age stars, as is the case for DDO
190, produces metallicities systematically lower than the real one. As a
consequence, the metallicity given above must be understood as a lower limit
and the quoted errors as mere estimates of the error of the mean, calculated
from the internal dispersion of the used sample. In any
case, the uncertainty does not significantly affect the resulting distance.

$(V-I)_{\rm TRGB}$ and $(V-I)_{-3.5}$ can finally be introduced in the
relationships of Lee  et al. (1993) to obtain the absolute magnitude of
the TRGB, $M_{I,\rm TRGB}=-4.00$, and hence a distance modulus of
$(m-M)_0=27.3\pm 0.2$, which corresponds to a distance to the Milky Way of
$d_{\rm MW}=2.9\pm 0.2$ Mpc.

The distance of DDO 190 to the barycenter of the LG and to other nearby
galaxy groups can be estimated. The position of the LG barycenter can be
calculated neglecting the masses of all the galaxies in the LG except Andromeda
and the Milky Way; assuming that the mass of the Milky Way is 0.7 times that
of Andromeda (Peebles 1989) and adopting a value of 0.77 Mpc for the Milky
Way--Andromeda distance (Freedman \& Madore 1990). The resulting distance for
DDO 190 to the barycenter of the LG is $2.9\pm 0.2$ Mpc.

The distances of DDO 190 to the closest groups of galaxies are 2.4 Mpc to the
M 81 group and 2.1 Mpc to the M 94 group (CnV-I). These values have been
obtained assuming that distances from the Milky Way are 3.6 Mpc for M 81
(Freedman et al. 1994) and 4.7 Mpc for M94 (from the velocity of M94,
$v=330$kms$^{-1}$, with $H_0=70$ km s$^{-1}$Mpc$^{-1}$). Angular distances
are $41^\circ.1$ from DDO 190 to M 81 and $17^\circ.5$ from DDO 190 to M94.

The closest known object to DDO 190 seems to be DDO 187. Both galaxies are
similar and are separated by some $21^\circ.5$ on the sky. Using the former
distance for DDO 190 and 2.4 Mpc for DDO 187 (Aparicio et al.  2000), the
distance from one galaxy to another is about 1.0 Mpc. Their radial velocities
relative to the barycenter of the Local Group are 275
km s$^{-1}$ for DDO 190 (average of the values by Fisher \& Tully 1981 and
Huchtmeier \& Richter 1988) and 136 km s$^{-1}$ for DDO 187 (Kraan-Korteweg
\& Tammann 1979). Hence the possibility that both galaxies could form a
physical pair can be rejected.

Summarizing, DDO 190 seems to be an isolated field galaxy, at more than 2 Mpc
from any galaxy group or giant galaxy and 1 Mpc from the nearest isolated
dwarf galaxy.

\section{Ionized gas}

\begin{figure}
\centerline{\psfig{figure=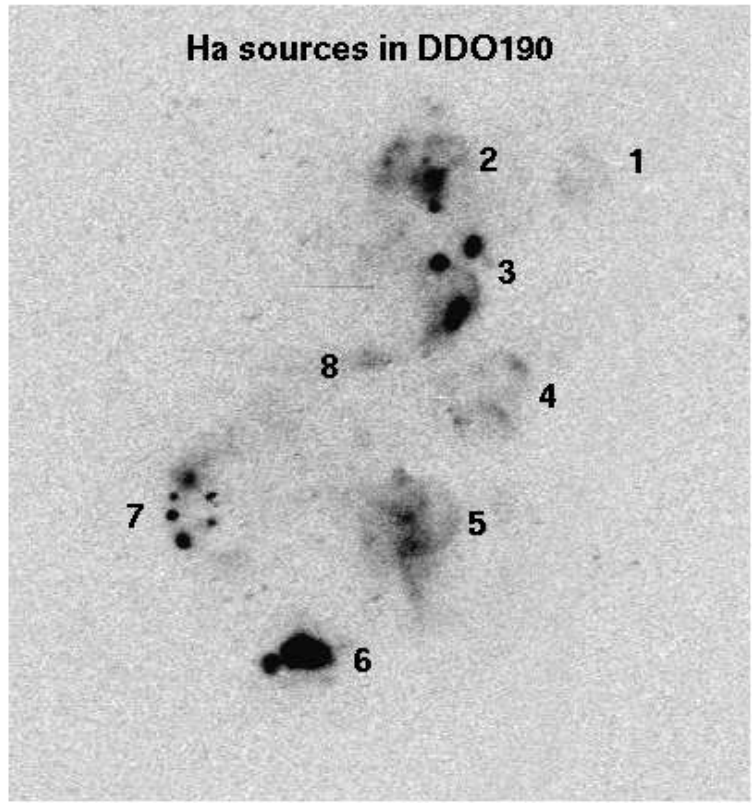,width=16cm}}
\figcaption[ha_orientada.eps]{H$_\alpha$ image of DDO 190. The total field is
$88\times 94 ('')^2$ and the integration time 900 s. North is up, east is
to the left.
\label{ha_orientada}}
\end{figure}

\placefigure{ha_orientada}
\placetable{fluxes}

Figure \ref{ha_orientada} shows the H$_\alpha$ image of DDO 190, after
subtraction of the continuum. We have identified eight H~{\sc ii} regions of different
morphology and compactness. A summary of their properties is given in Table
\ref{fluxes}. Column 1 gives the identification number as in
Figure \ref{ha_orientada}. Column 2 gives the flux received from each
region. Column 3 gives the corresponding H$_\alpha$ luminosities calculated
for a distance of 2.9 Mpc. Column 4 lists the number of ionizing photons per
second required to produce those luminosities, calculated following Kennicut
(1988) and assuming no extinction, which obviously provides a lower limit to
the number of emitted photons. Using also data by Kennicut (1988), the upper
limit would be about twice the quoted values.

The last line in Table \ref{fluxes} gives quantities integrated for the whole
galaxy, including also the very diffuse emission. The current SFR can be
estimated from $N_L$. The relations given in Aparicio et al. (2000) have been
used. The reader is referred to this paper for more details. $N_L$ is given
by
\begin{equation}
N_L=\int_{m_k}^{m_s}\int_{0}^{\tau(m)}\phi(m)\,\psi_n(t)\,n_{\rm L}(m)\,dm\,dt,
\end{equation}
\noindent where time increases from the present to the past, the present time
being 0 (i. e. it is the age); $\phi(m)$ is the IMF; $\psi_n(t)$ is the SFR
in units of number of stars per year; $\tau(m)$ is the MS lifetime of a star
of mass $m$, and $n_{\rm L}(m)$ is the number of Lyman photons emitted per
second by an MS star of mass $m$. While we are interested in stars more
massive than some 10 M$_{\sun}$ (see Aparicio et al.  2000) we need to
consider only the interval of time 0 to $\tau(10)$. Assuming, for simplicity,
that the SFR is constant over this interval and using the relations given in
Aparicio et al.  (2000) with the IMF of Kroupa, Tout, \& Gilmore (1993),
equation (3) becomes
\begin{equation}
N_{\rm L}=S\times \psi_n(0),
\end{equation}
\noindent with $S=1.29\times 10^{52}$, $2.41\times 10^{52}$ and $3.34\times
10^{52}$ for $m_s=30$M$_{\sun}$, $50$M$_{\sun}$ and $80$M$_{\sun}$,
respectively. The result is almost insensitive to the value of $m_k$, which
has been fixed to $m_k=12$M$_{\sun}$. From this, the current SFR results
$\psi_n(0)=(2.7^{+2.7}_{-1.3})\times 10^{-2}$yr$^{-1}$ or
$\psi(0)=(1.4^{+1.4}_{-0.7})\times 10^{-2}$M$_{\sun}$yr$^{-1}$. The latter
has been obtained using an average stellar mass 0.5 M$_{\sun}$, which
results from integration of the Kroupa et al.'s IMF. The errors have been
adopted to confidently account for the uncertainty introduced by the choice of
the upper limit for the stellar mass, as well as the uncertainty of the
internal extinction of the HII regions.

\section{Morphology and radial distribution of stellar populations}

Evident galactocentric gradients in the stellar populations have been
recently found in dwarf galaxies (Aparicio et al.  2000; Schulte-Ladbeck
et al. 1999; Mart\'\i nez-Delgado, Gallart, \& Aparicio 1999; Minniti,
Zijlstra, \& Alonso 1999; Aparicio et al.  1997b; Minniti \& Zijlstra
1996). They indicate that young stars are much more concentrated than
intermediate and old stars. This could be the result of the existence of old
extended halos in these galaxies. As we will show later on, we have not
covered a large enough field for DDO 190 to determine the total extension of
the galaxy. However, the radial-density profile can still be studied and some
conclusions can be traced out regarding the different extensions of the young
and the intermediate-to-old stars.

\begin{figure}
\centerline{\psfig{figure=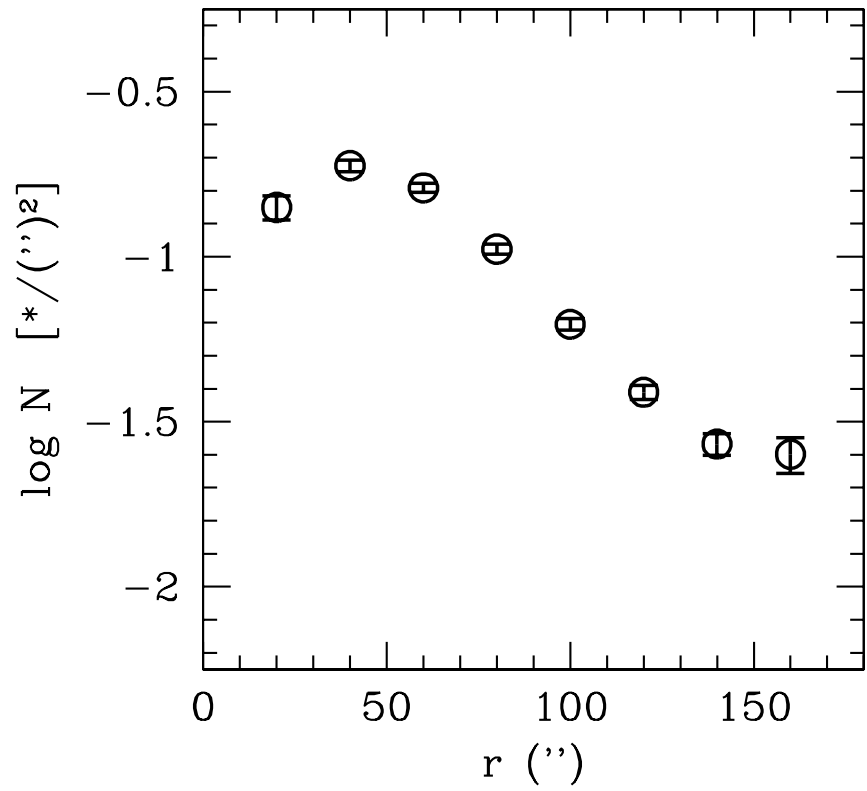,width=16cm}}
\figcaption[densi.eps]{Resolved star density profile of DDO190. Semi-major
axis of the ellipses quoted in the text are represented in the horizontal
axis.
\label{densi}}
\end{figure}

\placefigure{densi}

We have searched the morphology of the distribution of resolved stars and
fitted it with ellipses. To determine the center, position angle, and
ellipticity of the isophotes IRAF was used to create an empty image to which
artificial stars were added with the magnitudes and positions of the stars
resolved and measured in DDO 190. This image was filtered with a $\sigma=50$
pixels Gaussian and IRAF was used to fit elliptical isophotes to it. These
ellipses were used afterwards to determine the stellar radial density
profile, which is shown in Figure \ref{densi}.

The innermost ellipses are clearly affected by incompleteness (see
Sect. 7). More interestingly, the number density of stars in the outermost
ellipse is $2.5\times 10^{-2} ('')^{-2}$. The number density of stars in the
comparison field is $3.2\times 10^{-3} ('')^{-2}$, indicating that we are far
from having reached the radius at which the galaxy completely vanishes. This
fact impedes deriving the integrated magnitude of the galaxy, but the scale
length can still be obtained from the stellar density profile. Using the
region $50''\leq r\leq 130''$, it results $\alpha=43\farcs4$, equivalent to
611 pc. For latter calculations and for comparison with other dIrr galaxies,
it is useful to estimate the Holmberg radius $r^B_{26.5}$. Using the former
value of $\alpha$ and $r^B_{25.0}=2\farcm0$ from Fisher, \& Tully (1981), it
results $r^B_{26.5}=3\farcm0$.

\begin{figure}
\centerline{\psfig{figure=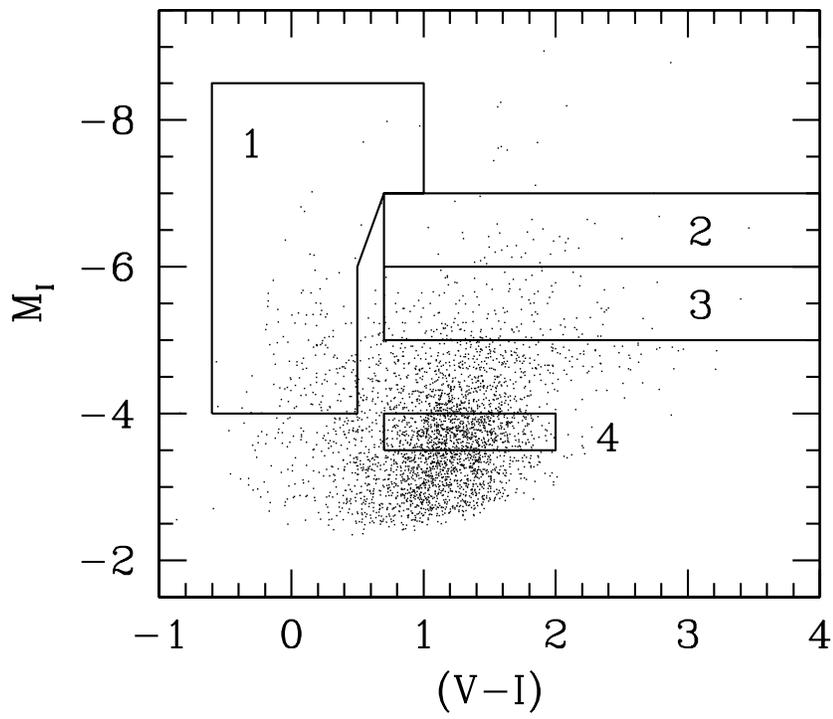,width=16cm}}
\figcaption[cmd_reg.eps]{CMD of DDO190 (same as shown in
Figure   \protect\ref{cmd}) showing the four regions used to parameterize the
distribution of stars of different ages. A distance modulus of
$(m-M)_)=27.3$ has been used. See also Fig. \protect\ref{cmd_par}.
\label{cmd_reg}}
\end{figure}

\begin{figure}
\centerline{\psfig{figure=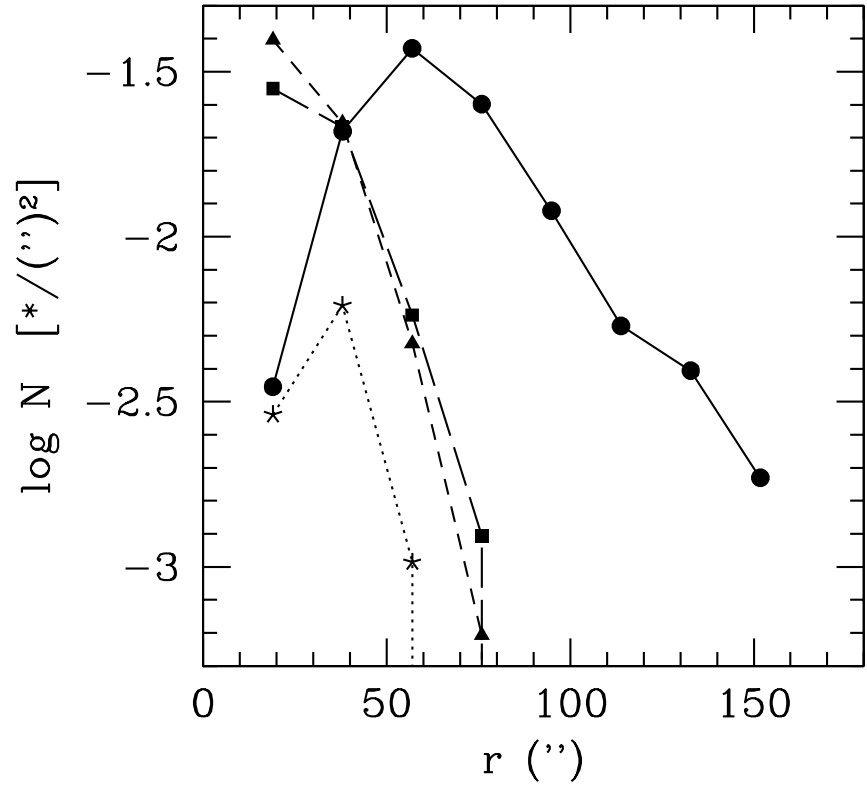,width=16cm}}
\figcaption[densi.eps]{Resolved star density profile of four stellar
populations of different ages in DDO 190. Stars have been selected from the
boxes shown in Figure   \protect\ref{cmd_reg}. Different line and point types
correspond to different boxes as follows. Dotted line, stars: box 1;
short-dashed line, triangles: box 2; long-dashed line, squares: box 3; full
line, dots: box 4.
\label{densi_pop}}
\end{figure}

\placefigure{cmd_reg}
\placefigure{densi_pop}

Whether different stellar populations are distributed differently in the
galaxy can be checked by studying the density profile of stars of different
ages. To do so, stars of four different regions of the CMD have been
selected. These boxes are shown in Figure \ref{cmd_reg}. The region between
boxes 3 and 4 has been excluded because it may be seriously affected by
observational effects; in particular by up-ward migration of RGB stars
(Aparicio, \& Gallart 1995; Mart\'\i nez-Delgado, \& Aparicio 1997). 

A synthetic CMD computed upon the Padua stellar evolutionary library
(Bertelli et al.  1994), has been used to define the four regions and to
check the ages sampled by each one. In particular, box 1 is populated by
stars younger than 0.1 Gyr only. Box 2 contains stars younger than 1 Gyr. Box
3 samples the age interval from 1 to 4 Gyr, preferentially. Finally, box 4 is
the only one containing a large amount of stars older than 4 Gyr (see \S 7
and Fig. \ref{cmd_par}, which illustrates this point). More details about the
distribution of stars as a function of age can be seen in Aparicio (1997).

The profiles for each region are shown in Figure
\ref{densi_pop}. Crowding effects in the innermost region of the galaxy
produce a decrease in the stellar density in the inner regions. Apart from
this, the figure shows that the youngest population completely vanishes at
less than $1'$ from the center, while the density of older stars (box 4) goes
down slowly. The intermediate populations show an intermediate behavior, but
it is interesting to note that at more than some $80''$ the only surviving
population is that of box 4. This can be interpreted as only stars older than
4 Gyr populating the outermost regions of the galaxy. This will be further
discussed in \S 7, where the SFR is calculated for increasing galactocentric
distances.

\section{The star formation history of DDO 190}

\begin{figure}
\centerline{\psfig{figure=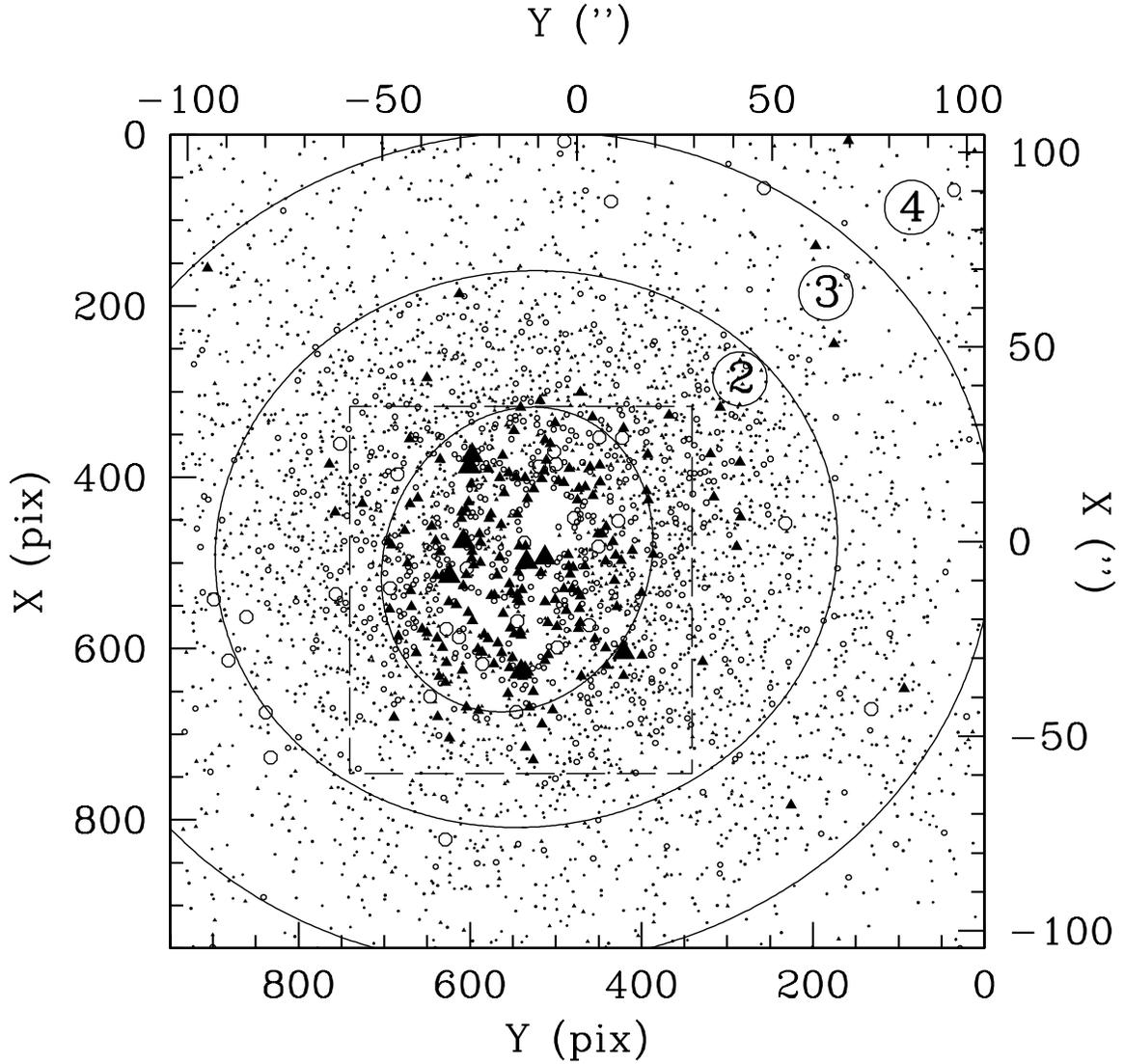,width=16cm}}
\figcaption[mapa.eps]{Resolved stars in DDO 190. Red stars ($V-I>0.8$) are
represented by open circles. Blue stars ($V-I\leq 0.8$) are represented by
filled triangles. The four regions in which the galaxy has been divided for
the stellar population distribution study are shown by the over-plotted
ellipses. Region 1 is the region enclosed by the innermost ellipse and has
not been labeled for clarity. Dashed line approximately corresponds to the
area covered by Figure \protect\ref{ha_orientada}. North is up and east is to
the left.
\label{mapa}}
\end{figure}

The SFH can be derived in detail for the full history of a galaxy from a deep CMD through
comparison with synthetic CMDs (see Gallart et al. 1999). The CMD of
DDO 190 is not deep enough for such a detailed analysis but it is still
possible to sketch the SFH of the galaxy for old, intermediate and young ages
and to study the spatial extension of each age interval. For this, the galaxy
has been divided into four elliptical regions of increasing size. The regions
are shown in Figure \ref{mapa} and 
have been selected according to the results of the elliptical isophote
analysis explained in \S 6. It must be noted that the small areas devoided of
resolved stars in the innermost regions correspond to the most luminous HII
regions discussed in \S 5. 

\begin{figure}
\centerline{\psfig{figure=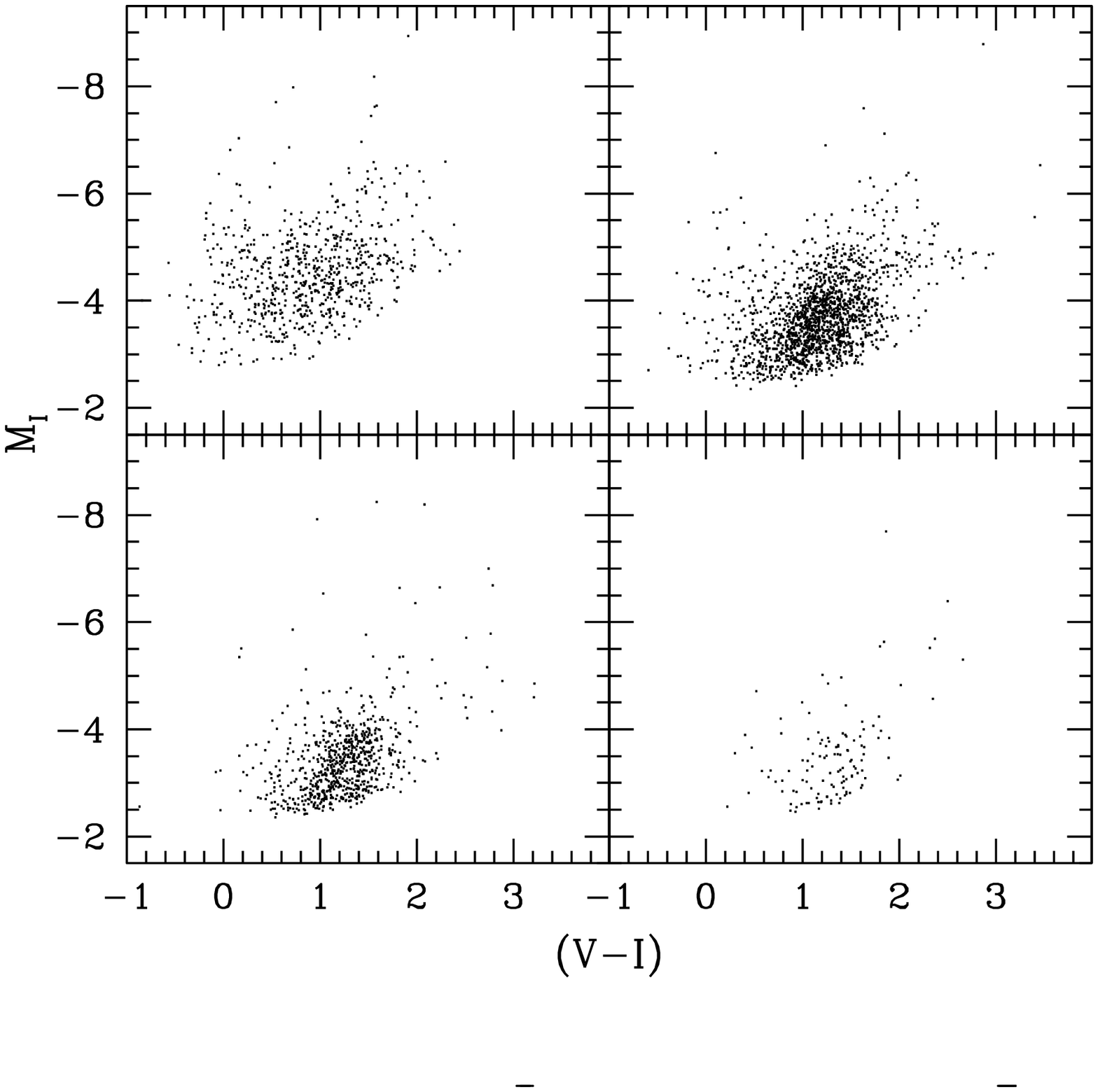,width=16cm}}
\figcaption[cmd_eli.eps]{CMDs of DDO190 for the four elliptical regions shown
in Figure  \protect\ref{mapa}. A distance modulus of
$(m-M)_)=27.3$ has been used. 
\label{cmd_eli}}
\end{figure}

The CMDs of the four elliptical regions are shown in Figure \ref{cmd_eli}. In
good agreement with the results obtained in \S 6, two main facts are apparent
from these figures: first that the young, blue stars, are present only in the
inner regions; and second that in the outer region, traces of the RGB are
still apparent.

\placefigure{mapa}
\placefigure{cmd_eli}

In the following we will quantitatively derive the stellar age distribution
in each elliptical region of DDO 190. Rigorously, this represents the SFH of
each region only if no migration of stars from one region to another has
occurred along their lives. For simplicity, and because we are not now
interested in the kinematical evolution of the galaxy, we will assume that
this is the case and we will call the resulting distributions the SFH. But it
must be kept in mind that what we are going to derive are current spatial and
age distributions of stars.

To derive the SFH, the same procedure used by Aparicio et al. (2000) has been
applied. It is a simplified version of the method proposed by Aparicio,
Gallart and Bertelli (1997a). In practice, a synthetic CMD with constant SFR
of value $\psi_0$, the IMF by Kroupa et al. (1993),  25 \% of binary
stars with mass ratios, $q$, uniformly distributed in the interval $0.6<q<1.0$
and a metallicity $Z$ taking random values from $Z_1=0.0002$ to $Z_2=0.001$,
independent of age have been used. The $Z$ interval has been chosen
considering the results found for the metallicity in \S 4.

\begin{figure}
\centerline{\psfig{figure=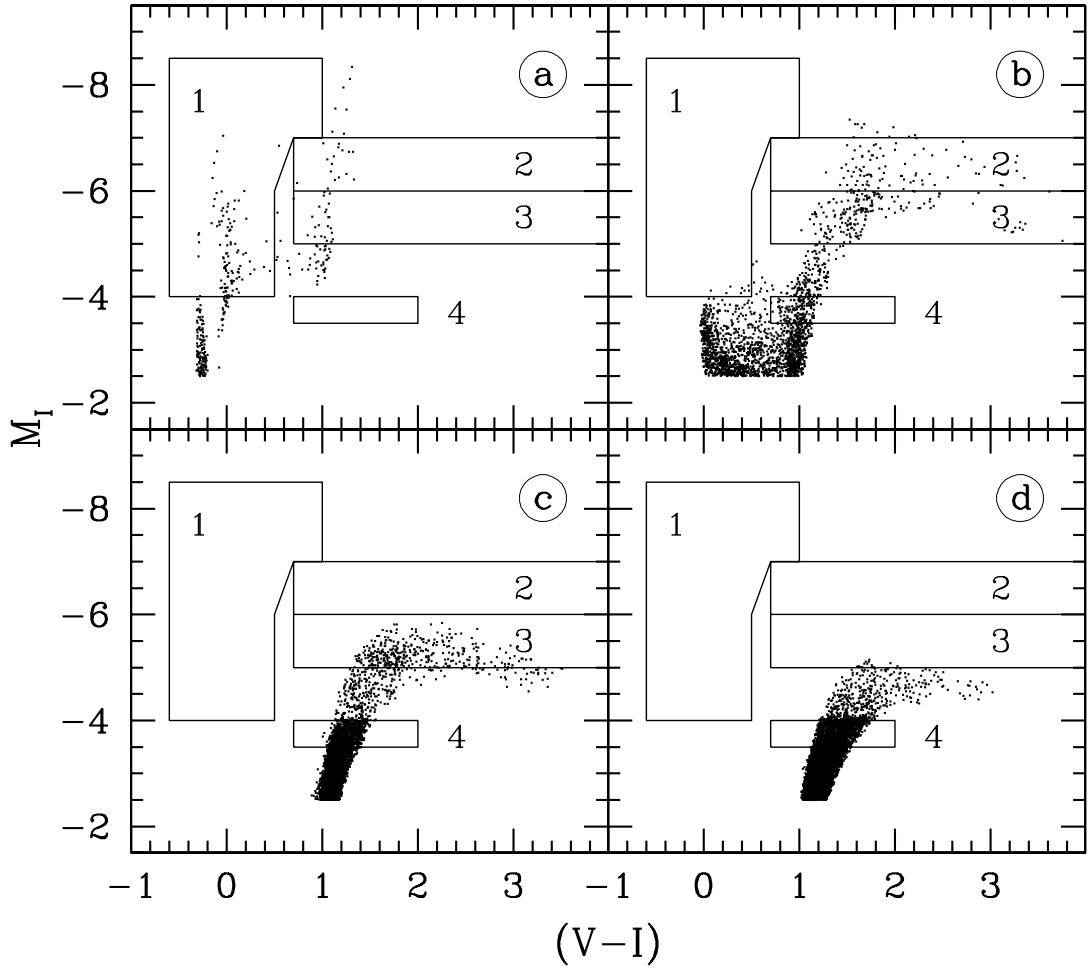,width=16cm}}
\figcaption[cmd_par.eps]{The adopted synthetic, partial CMDs. The
corresponding age intervals are: a) 0--0.1 Gyr; b) 0.1--1 Gyr; c) 1--4 Gyr;
d) 4--15 Gyr. The four boxes defined to sample the stellar populations (see
\S 6 and Fig. \protect\ref{cmd_reg}) are
overplotted.
\label{cmd_par}}
\end{figure}

\placefigure{cmd_par}

The resulting synthetic CMD has then been divided into four age intervals:
0--0.1 Gyr, 0.1--1 Gyr, 1--4 Gyr, and 4--15 Gyr. Following the nomenclature
introduced in Aparicio et al.  (1997a), each of these synthetic
diagrams will be called {\it partial model} CMD and any linear combination of
them will be denoted as a {\it global model}. The four partial model CMDs are
shown in Fig. \ref{cmd_par}.

For each elliptical region, stars have been counted in the
four boxes shown in Figures \ref{cmd_reg} and \ref{cmd_par}. We will denote
by $N_j^o$ the number of stars of the observed CMD lying in region $j$ and
by $N_{ji}^m$ the number of stars of partial model (or age interval) $i$
populating region $j$. The number of stars populating a given region in a
global model is then given by
\begin{equation}
N_j^m=k\sum_i\alpha_iN_{ji}^m
\end{equation}
\noindent and the corresponding SFR by 
\begin{equation}
\psi(t)=k\sum_i\alpha_i\psi_0\Delta_i(t),
\end{equation}
\noindent where $\alpha_i$ are the linear combination coefficients, $k$ is a
normalization constant, and $\Delta_i(t)=1$ if $t$ is inside the interval
corresponding to partial model $i$ and $\Delta_i(t)=0$ otherwise. The
$\psi(t)$ having the best compatibility with the data can be obtained by a
least-squares fit of $N_j^m$ to $N_j^o$.

The quantities $N_j^o$ must be corrected of incompleteness. Completeness
factors have been estimated for each elliptical region shown in
Figure \ref{mapa} and for the four boxes defined in the CMD. The usual
artificial star experiments and DAOPHOT have been used (Stetson 1987). Two
arrays of artificial stars have been builded up with stars separated 30
pixels in $x$ and $y$ coordinates from each other. In this way, it is
guaranteed that no over-crowding (crowding of artificial stars between each
other) is produced. Magnitudes and colors have been chosen to sample the
magnitudes and colors of the boxes 1 and 2 for the first array and 3 and 4
for the second array. In practice $-4\geq M_I\geq -6$ and $(V-I)=0.2$ has
been used to sample box 1; $M_I=-6.5$, $(V-I)=1.8$ for box 2; $M_I=-5.5$,
$(V-I)=1.6$ for box 3, and $M_I=-3.75$, $(V-I)=1.3$ for box 4. Each array has
been added to the $V$ and $I$ images creating two sets of images and the
photometry has been recalculated for each set in the same way as for the real,
original images. By this procedure, some 50 artificial stars are added in
each CMD box to the smallest (the innermost) ellipse and more than 500 in
each CMD box for the whole image.

The completeness factors for each box and ellipse are then calculated as the
rates of the recovered to injected artificial stars. The criteria to
determine when an artificial star is recovered are the same discussed in
Aparicio, \& Gallart (1995). Completeness factors are 1.0 for all the boxes
in all the elliptical regions, except the innermost one. In this,
completeness is 1.0 for box 2; 0.85 for boxes 1 and 3 and 0.25 for box
4. These values have been used to correct the star counts in the
observational CMD, but the low completeness in box 4 of ellipse 1 introduces
a greater uncertainty in the results for the oldest population in the
innermost region of the galaxy.

\begin{figure}
\centerline{\psfig{figure=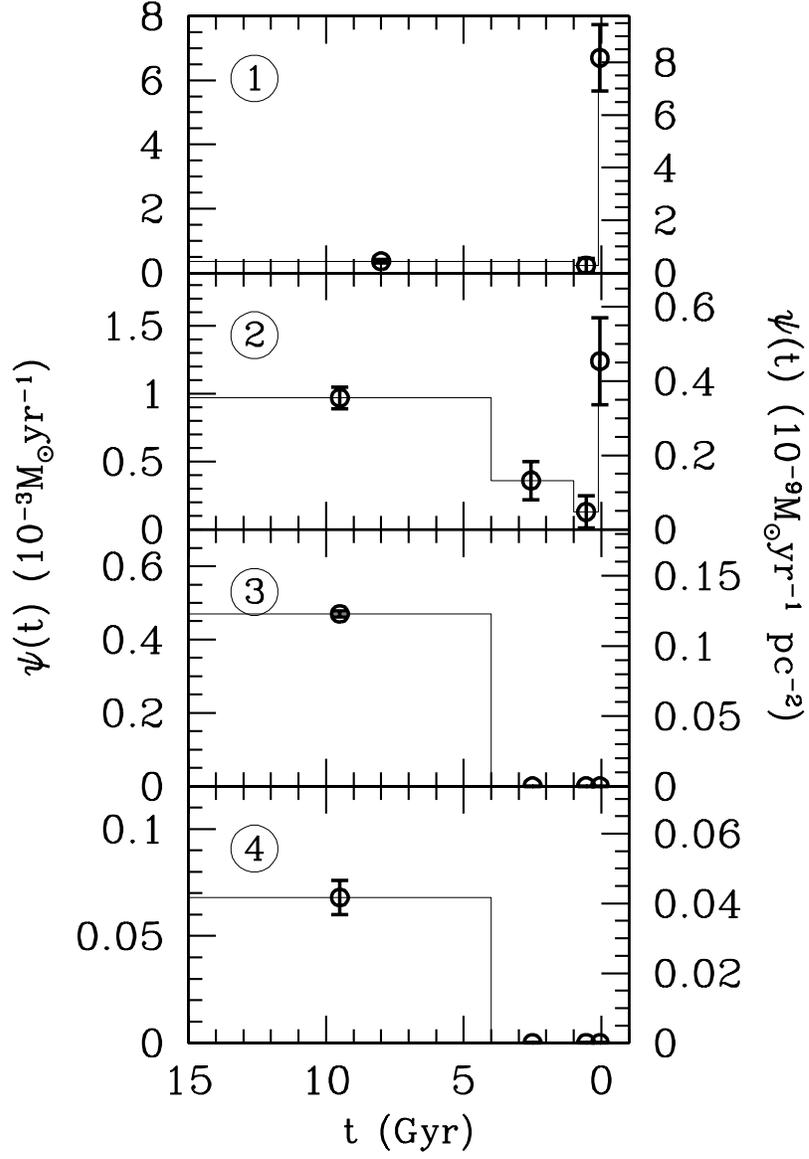,width=16cm}}
\figcaption[psi.eps]{The SFR of DDO 190, obtained from the analysis of
its CMD using synthetic CMDs. Panels correspond to the four elliptical
regions shown in Figure \protect\ref{mapa}.
\label{psi}}
\end{figure}

\begin{figure}
\centerline{\psfig{figure=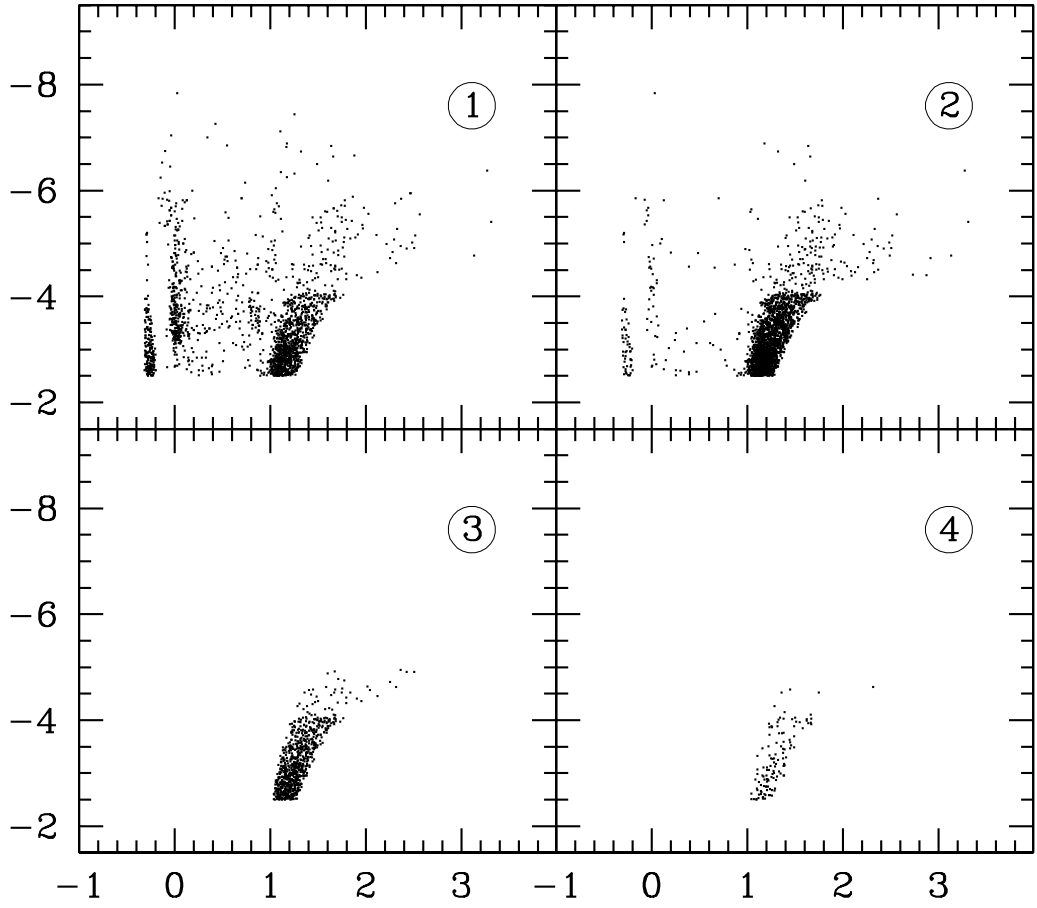,width=16cm}}
\figcaption[psi.eps]{Global sinthetic CMDs corresponding to the best, adopted
solutions for $\psi(t)$ shown in Figure \protect\ref{psi}. Panels correspond
to the four elliptical regions shown in Figure \protect\ref{mapa}.
\label{cmd_sin}}
\end{figure}

\placefigure{psi}
\placetable{sfr}
\placefigure{cmd_sin}

The resulting $\psi(t)$ for DDO 190 is shown in Figure \ref{psi}. The
right-hand vertical scale is normalized to the area covered by each
region. Note that for region 1 the current SFR estimated from the H$_\alpha$
luminosity is more than twice as large as that estimated from the CMD but
error bars of both marginally overlap. The former is not represented in
Figure \ref{psi}.

Crowding is more severe in the innermost region of the galaxy. This makes the
estimate of $\psi(t)$ for intermediate and old stars more uncertain. For this
reason, we prefer giving an average for the 1 to 15 Gyr interval (see panel
1 of Fig. \ref{psi}).

Global model CMDs corresponding to the adopted
solutions for $\psi(t)$ are shown in Figure \ref{cmd_sin}. This figure is
intended as only ilustrative. It particular, when compared with the
observational CMDs shown in Figure \ref{cmd_eli} it must be kept in mind
that, in Figure \ref{cmd_sin} the observational effects have not been
simulated, resulting in the most crowding affected region to be overpopulated
in this figure.

A summary of the SFR averaged over different intervals of time is given in
Table \ref{sfr}. Column 1 indicates the interval of time and whether the SFR
has or has not been normalized to the area. The interval of time used is
given in Gyr by subscripts of the averaged SFR $\bar\psi$ (e.g.,
$\bar\psi_{4-1}$ stands for the SFR averaged for the 4 to 1 Gyr ago period of
time). The first seven lines list values for $\bar\psi$ not normalized to the
area.  The present-day value, derived from the H$_\alpha$ luminosity, is
denoted by $\psi(0)$. Lines 8 to 14 give the same normalized to the area
covered by each region. Columns labeled 1 to 4 give the data for the four
studied regions. Last column gives extrapolated estimates for the outermost
region of the galaxy, not covered by our image, down to the Holmberg radius
$r^B_{26.5}$. These values have been obtained by extrapolation of the
quantities corresponding to region 4 and using an exponential function of
scale length $\alpha=43.''4$, as found above for the stellar density profile
of the galaxy. It must be stressed that these values are only crude
estimates.

The analysis of Figure \ref{psi} and Table \ref{sfr} shows that while the
older population follows a smooth decreasing density radial distribution,
young stars are only present in the inner regions. In particular, a
negligible number of stars younger than 4 Gyr appears in the three
outermost regions, i.e., at galactocentric distances greater than $80''$ or
1.10 kpc (semi-major axis) while almost all the stars younger than 0.1 Gyr
are concentrated in the innermost ellipse, of $40''$ or 0.55 kpc. The effect
is even stronger if the SFR estimated from the H$_\alpha$ luminosity is
considered, as it is concentrated in region 1.

With different levels of evidence, the behavior of the stellar populations of
DDO 190 is being found in all the dIrr galaxies for which a spatially
extended study has been performed (see the discussion in Aparicio et al.
2000). This could possibly be related to the kinematical history of the
galaxy and the oldest stars could be tracing its primeval structure and the
possible dark matter potential well. None of the results found till now
refutes this idea. However, the final confirmation should ideally come from
more observational data including, in particular, the kinematics, which is,
unfortunately, still beyond the reach of currently available instruments.

\section{Global properties of DDO 190}

\placetable{global}

The SFR $\psi(t)$ and the new estimate of the distance obtained here have
been used, together with data from other authors, to calculate the integrated
properties of DDO 190 listed in Table \ref{global}. The bracketed figures
refer to the bibliographic sources for each datum. Nothing is indicated for
quantities directly derived from other quantities listed in the table. First
the coordinates, the metallicity, [Fe/H], and the distance to the Milky Way,
$d_{\rm MW}$, and to the barycenter of the Local Group, $d_{\rm LG}$, and the
semi-major axis to the Holmberg radius at $\mu_B=26.5('')^{-2}$ are
given. Integrated absolute magnitudes and luminosities referred to the
Holmberg radius at $\mu_B=25.0('')^{-2}$ follow. $M_\star$ is the mass in
stars and stellar remnants for the whole galaxy, calculated from integration
of $\psi(t)$ for the whole galaxy life and assuming that 0.8 of this integral
remains locked in stellar objects; $M_{\rm gas}$ is obtained by multiplying
the H~{\sc i} mass by 4/3 to take into account the mass in He; $M_{\rm vt}$
is the virial mass obtained from the velocity dispersion of the gas; $\mu$ is
the gas fraction, relative to the total mass intervening in the chemical
evolution (mainly gas and stellar objects; see Peimbert et al. 1993), and
$\kappa$ is the dark matter fraction, calculated as indicated in the
table. Finally, the gas and total mass to luminosity fractions are given.

\section{Conclusions}

We have presented $V$ and $I$ CCD photometry of $\sim3900$ stars in the dIrr
galaxy DDO 190, as well as H$_\alpha$ photometry of its H~{\sc ii}
regions. The color--magnitude diagram shows a red giant branch, and a
sequence of blue stars. From the mean color of the RGB, $(V-I)_{0,-3.5}=
1.27$, a mean metallicity [Fe/H] $=-2.0$ is obtained. Based on the $I$
magnitude of the TRGB we have derived the distance modulus of DDO 190 to be
$(m-M)_0=27.3\pm 0.2$, which corresponds to a distance from the Milky Way of
$d_{\rm MW}=2.9\pm 0.2$ Mpc. The distances of DDO 190 to the nearest galaxy
groups are 2.1 Mpc to the M 94 group (CnV-I), 2.4 Mpc to the M 81 group and
2.9 Mpc to the barycenter of the LG. The nearest known galaxy to DDO 190 is
DDO 187, at 1.0 Mpc, all this indicating that DDO 190 is an isolated field
galaxy.

The surface-brightness distribution of the galaxy is well fitted by ellipses
of ellipticity $e=1-a/b=0.1$ and $p.a.=82^\circ$. These parameters have been
used to find the radial star density distribution, which is fitted by an
exponential law of scale length $\alpha=43\farcs4$, corresponding to 611
pc. The galaxy is bigger than the field covered by our images. The former
value of $\alpha$ and previously published surface photometry have been used
to estimate a value for the Holmberg semi-major axis to $\mu_B=26.5$ of
$r^B_{26.5}=3\farcm0$.

The star-formation history of DDO 190 has been estimated from the number of
blue and red stars in the color--magnitude diagram, using a synthetic CMD
as reference. The current SFR has also been estimated from the H$_\alpha$
luminosity. The star-formation rate has been evaluated for the intervals of
time 0--0.1 Gyr, 0.1--1 Gyr, 1--4 Gyr, and 4--15 Gyr, for which the values
summarized in Table \ref{sfr} have been obtained.

Stellar populations of different ages show different spatial distributions,
the oldest being much more extended than the youngest. In fact DDO 190 shows
a current strong enhancement in the SFR in its central region, which is a
common feature of galaxies classified as dIrr. Stars younger than 0.1 Gyr
occupy only the central $40''$ (0.55 kpc); stars younger than a few ($\sim
4$) Gyr extend up to $\sim 80''$ (1.10 kpc), and for larger galactocentric
distances only older stars seem to be present. This behavior is being found
in all the dIrr galaxies for which spatially extended studies have been
performed and could be related to the kinematical history of the galaxy. In
particular the oldest stars could being tracing the primeval structure of the
galaxy and its possible dark-matter potential well (see a thorough discussion
of this in Aparicio et al. 2000 and references therein). This, however, needs
stronger observational confirmation, kinematical data being meant to provide
very useful information.

\acknowledgements

AA is financially supported by the IAC (grant P3/94), by the DGESIC of the
Kingdom of Spain (grant PB97-1438-C02-01) and by the DGUI of the Autonomous
Government of the Canary Islands (grant P99-008). NASA's Extragalactic Database (NED) has been
used in our work.

\newpage

\begin{deluxetable}{cccccc}
\tablenum{1}
\tablewidth{0pt}
\tablecaption{Journal of observations
\label{journal}}
\tablehead{
\colhead{Date} & \colhead{Object} & \colhead{Time (UT)} & \colhead{Filter} &
\colhead{Exp. time (s)} & \colhead{FWHM ($''$)}}
\startdata
97.07.28 & DDO 190 & 21:03 & $V$ & 1200 & 0.60 \nl
97.07.28 & DDO 190 & 21:24 & $V$ & 1200 & 0.60 \nl
97.07.28 & DDO 190 & 21:45 & $I$ & 1000 & 0.50 \nl
97.07.28 & DDO 190 & 22:03 & $I$ & 1000 & 0.55 \nl
97.07.29 & DDO 190 & 21:35 & H$_\alpha$ & 900 & 0.75 \nl
97.07.29 & DDO 190 & 21:50 & H$_\alpha$ & 900 & 0.75 \nl
97.07.29 & DDO 190 & 22:06 & H$_\alpha$-cont & 600 & 0.80 \nl
97.07.29 & Field & 23:08 & $I$ & 600 & 1.05 \nl
97.07.29 & Field & 23:18 & $I$ & 600 & 0.80 \nl
97.07.29 & Field & 23:30 & $V$ & 600 & 0.80 \nl
97.07.29 & Field & 23:40 & $V$ & 600 & 0.95 \nl
\enddata

\end{deluxetable}

\newpage
\begin{deluxetable}{cccc}
\tablenum{2}
\tablewidth{0pt}
\tablecaption{H~{\sc ii} regions in DDO 190
\label{fluxes}}
\tablehead{
\colhead{Id.} & \colhead{Flux} &
\colhead{Luminosity} & \colhead{$N_L$} \nl
\colhead{~} & \colhead{$10^{-13}$erg\,s$^{-1}$\,cm$^{-2}$} &
\colhead{$10^{38}$erg\,s$^{-1}$} & \colhead{$10^{49}$s$^{-1}$}}
\startdata
1 & 0.12 & 0.12 & 0.88 \nl
2 & 1.04 & 1.01 & 7.38 \nl
3 & 1.29 & 1.25 & 9.12 \nl
4 & 0.41 & 0.40 & 2.92 \nl
5 & 1.21 & 1.17 & 8.54 \nl
6 & 2.72 & 2.63 & 19.20 \nl
7 & 0.90 & 0.87 & 6.35 \nl
8 & 0.15 & 0.15 & 1.10 \nl
Total & 9.34 & 9.05 & 66.06 \nl
\enddata

\end{deluxetable}

\begin{deluxetable}{lccccc}
\tablenum{3}
\tablewidth{420pt}
\tablecaption{Summary of the Star Formation History of DDO 190
\label{sfr}}
\tablehead{\colhead{} & \multicolumn{5}{c}{Regions} \nl
\colhead{$\bar\psi$} & \colhead{1} &
\colhead{2} & \colhead{3} & \colhead{4} & \colhead{Outer}} 
\scriptsize
\startdata 

$\bar\psi_{15-0}$\hfill ($10^{-3}$M$_{\sun}$yr$^{-1}$) \hspace {8mm} &
$0.38\pm 0.08$ & $0.80\pm 0.04$ & $0.350\pm0.006$ & $0.050\pm 0.006$ & 
0.1: \nl 
$\bar\psi_{15-1}$ \hfill ($10^{-3}$M$_{\sun}$yr$^{-1}$)
\hspace {8mm} & $0.36\pm 0.05$ & $0.84\pm 0.04$ & $0.368\pm 0.006$ & 
$0.053\pm 0.006$ & 0.1: \nl 
$\bar\psi_{15-4}$ \hfill ($10^{-3}$M$_{\sun}$yr$^{-1}$)
\hspace {8mm} & \nodata & $0.97\pm 0.09$ & $0.470\pm 0.008$ & 
$0.068\pm 0.008$ & 0.15: \nl 
$\bar\psi_{4-1}$ \hfill ($10^{-3}$M$_{\sun}$yr$^{-1}$)
\hspace {8mm} & \nodata & $0.36\pm 0.14$ & 0 & 0 & 0 \nl 
$\bar\psi_{1-0}$ \hfill ($10^{-3}$M$_{\sun}$yr$^{-1}$) \hspace {8mm} &
$0.88\pm 0.13$ & $0.24\pm 0.10$ & 0 & 0 & 0 \nl 
$\bar\psi_{0.1-0}$ \hfill ($10^{-3}$M$_{\sun}$yr$^{-1}$) \hspace {8mm} &
$6.7\pm 1.0$ & $1.2\pm 0.3$ & 0 & 0 & 0 \nl 
$\psi(0)$ \hfill ($10^{-3}$M$_{\sun}$yr$^{-1}$) \hspace {8mm} &
$14^{+14}_{-7}$ & 0 & 0 & 0 & 0 \nl 
$\bar\psi_{15-0}/A$ \hfill ($10^{-9}$M$_{\sun}$yr$^{-1}$pc$^{-2}$)
\hspace {8mm} & $0.46\pm 0.09$ & $0.29\pm 0.02$ & $0.092\pm 0.002$ 
& $0.030\pm 0.003$ & 0.01: \nl 
$\bar\psi_{15-1}/A$ \hfill ($10^{-9}$M$_{\sun}$yr$^{-1}$pc$^{-2}$)
\hspace {8mm} & $0.44\pm 0.06$ & $0.30\pm 0.02$ & $0.097\pm 0.002$ 
& $0.033\pm 0.004$ & 0.01: \nl 
$\bar\psi_{15-4}/A$ \hfill ($10^{-9}$M$_{\sun}$yr$^{-1}$pc$^{-2}$)
\hspace {8mm} & $0.44\pm 0.06$ & $0.36\pm 0.03$ & $0.123\pm 0.002$ 
& $0.042\pm 0.005$ & 0.015:\nl 
$\bar\psi_{4-1}/A$ \hfill ($10^{-9}$M$_{\sun}$yr$^{-1}$pc$^{-2}$)
\hspace {8mm} & $0.44\pm 0.06$ & $0.13\pm 0.05$ & 0 & 0 & 0 \nl 
$\bar\psi_{1-0}/A$ \hfill ($10^{-9}$M$_{\sun}$yr$^{-1}$pc$^{-2}$) \hspace
{8mm} & $1.1\pm 0.2$ & $0.09\pm 0.03$ & 0 & 0 & 0 \nl 
$\bar\psi_{0.1-0}/A$ \hfill ($10^{-9}$M$_{\sun}$yr$^{-1}$pc$^{-2}$) \hspace
{8mm} & $8.2\pm 1.3$ & $0.45\pm 0.12$ & 0 & 0 & 0 \nl 
$\psi(0)/A$ \hfill ($10^{-9}$M$_{\sun}$yr$^{-1}$pc$^{-2}$) \hspace {8mm} 
& $17^{+17}_{-8}$ & 0 & 0 & 0 & 0 \nl 

\enddata
\end{deluxetable}

\begin{deluxetable}{lc}
\tablenum{4}
\tablewidth{350pt}
\tablecaption{Global properties of DDO 190
\label{global}}
\tablehead{\colhead{} & \colhead{}}
\startdata 

$\alpha_{2000}$, $\delta_{2000}$ & $\rm 14^h 15^m 56^s$, $23^\circ 03' 13''$ \nl
$l, b$ & $25^\circ.6, +70^\circ.5$ \nl
$[Fe/H]$ & --2.0 (1) \nl 
$d_{\rm MW}$ \hfill (Mpc) & 2.9 (1) \nl 
$d_{\rm LG}$ \hfill (Mpc) & 2.9 (1) \nl 
$M_{B,0}$ & --14.30 (4) \nl 
$M_{V,0}$ & --14.59 (4) \nl
$r^B_{26.5}$ & 3\farcm0 (1,2) \nl
$L_B$ \hfill ($10^7$L$_\odot$) & 8.6 \nl 
$L_V$ \hfill ($10^7$L$_\odot$) & 6.1 \nl 
$M_\star$ \hfill ($10^7$M$_{\sun}$) & 1.8 (1) \nl 
$M_{\rm gas}$ \hfill ($10^7$M$_{\sun}$) & 6.6 (2,3) \nl 
$M_{\rm vt}$ \hfill ($10^7$M$_{\sun}$) & 29 (2,3) \nl
$\mu=M_{gas}/(M_\star+M_{gas})$ ~~~ & 0.79 \nl
$\kappa=1-(M_\star+M_{\rm gas})/M_{\rm vt}$ & 0.71 \nl 
$M_{\rm gas}/L_V$ \hfill (M$_{\sun}$/L$_\odot$) & 1.1 \nl 
$M_{\rm vt}/L_V$ \hfill (M$_{\sun}$/L$_\odot$) & 4.8 \nl 

\enddata

\tablerefs{(1) This paper; (2) Fischer, \& Tully (1981); (3) Hutchmeier, \&
Richter (1988); (4) Prugniel, \& H\'eraudeau (1998)}

\end{deluxetable}

\end{document}